\documentclass[conference]{IEEEtran}
\IEEEoverridecommandlockouts
\usepackage{cite}
\usepackage{amsmath,amssymb,amsfonts}
\usepackage{graphicx}
\usepackage{textcomp}
\usepackage{xcolor}
\usepackage{colortbl}
\usepackage{tcolorbox}
\usepackage[a4paper, total={184mm,239mm}]{geometry}
\usepackage{xspace}
\usepackage{booktabs}
\usepackage{url}
\def\BibTeX{{\rm B\kern-.05em{\sc i\kern-.025em b}\kern-.08em
    T\kern-.1667em\lower.7ex\hbox{E}\kern-.125emX}}

\usepackage{algorithm}
\usepackage{algpseudocode} 
\usepackage{lineno}
\usepackage{array}
\usepackage{makecell}

\newcommand{\highlight}[1]{\colorbox{lightgray}{#1}}

\usepackage{balance}
\usepackage{tabularx}

\usepackage{caption}
\usepackage{tikz}
\newcommand{\circled}[1]{%
  \tikz[baseline=(char.base)]{
    \node[shape=circle,draw,inner sep=0.5pt,fill=black,text=white] (char) {#1};}}
\newcommand{\tool}[1]{\textsc{#1}\xspace}
\newcommand{\toolname}{\tool{SaFliTe}}
\newcommand{\framename}{\tool{Universal Autononmous System Fuzzing with LLMs}}

\newcommand{\commentout}[1]{}

\begin{document}

\title{\toolname: Fuzzing Autonomous Systems via Large Language Models}

\author{Anonymous Author(s)}

\commentout{
\author{
    \IEEEauthorblockN{name1, name2, name3}
    \IEEEauthorblockA{The University of Manchester, UK\\
                      Email: \{name1, name2, name3\}@manchester.ac.uk}
}
}

\author{
    \IEEEauthorblockN{Taohong Zhu, Adrians Skapars, Fardeen Mackenzie, Declan Kehoe, William Newton}
    \IEEEauthorblockN{Suzanne Embury, Youcheng Sun}
    \IEEEauthorblockA{Department of Computer Science, University of Manchester, Manchester, United Kingdom}
    \IEEEauthorblockA{\{taohong.zhu/fardeen.mackenzie/declan.kehoe/suzanne.m.embury/youcheng.sun\}@manchester.ac.uk}
    \IEEEauthorblockA{\{adrians.skapars/william.newton\}@student.manchester.ac.uk}
}
\commentout{
\author{\IEEEauthorblockN{1\textsuperscript{st} Given Name Surname}
\IEEEauthorblockA{\textit{dept. name of organization (of Aff.)} \\
\textit{name of organization (of Aff.)}\\
City, Country \\
email address or ORCID}
\and
\IEEEauthorblockN{2\textsuperscript{nd} Given Name Surname}
\IEEEauthorblockA{\textit{dept. name of organization (of Aff.)} \\
\textit{name of organization (of Aff.)}\\
City, Country \\
email address or ORCID}
\and
\IEEEauthorblockN{3\textsuperscript{rd} Given Name Surname}
\IEEEauthorblockA{\textit{dept. name of organization (of Aff.)} \\
\textit{name of organization (of Aff.)}\\
City, Country \\
email address or ORCID}
\and
\IEEEauthorblockN{4\textsuperscript{th} Given Name Surname}
\IEEEauthorblockA{\textit{dept. name of organization (of Aff.)} \\
\textit{name of organization (of Aff.)}\\
City, Country \\
email address or ORCID}
\and
\IEEEauthorblockN{5\textsuperscript{th} Given Name Surname}
\IEEEauthorblockA{\textit{dept. name of organization (of Aff.)} \\
\textit{name of organization (of Aff.)}\\
City, Country \\
email address or ORCID}
\and
\IEEEauthorblockN{6\textsuperscript{th} Given Name Surname}
\IEEEauthorblockA{\textit{dept. name of organization (of Aff.)} \\
\textit{name of organization (of Aff.)}\\
City, Country \\
email address or ORCID}
}
}

\maketitle

\begin{abstract}
Fuzz testing effectively uncovers software vulnerabilities; however, it faces challenges with Autonomous Systems (AS) due to their vast search spaces and complex state spaces, which reflect the unpredictability and complexity of real-world environments.
This paper presents a universal framework aimed at improving the efficiency of fuzz testing for AS. At its core is \toolname, a predictive component that evaluates whether a test case meets predefined safety criteria. By leveraging the large language model (LLM) with information about the test objective and the AS state, \toolname assesses the relevance of each test case.

We evaluated \toolname by instantiating it with various LLMs, including GPT-3.5, Mistral-7B, and Llama2-7B, and integrating it into four fuzz testing tools: PGFuzz, DeepHyperion-UAV, CAMBA, and TUMB. These tools are designed specifically for testing autonomous drone control systems, such as ArduPilot, PX4, and PX4-Avoidance. The experimental results demonstrate that, compared to PGFuzz, \toolname increased the likelihood of selecting operations that triggered bug occurrences in each fuzzing iteration by an average of 93.1\%. Additionally, after integrating \toolname, the ability of DeepHyperion-UAV, CAMBA, and TUMB to generate test cases that caused system violations increased by 234.5\%, 33.3\%, and 17.8\%, respectively. The benchmark for this evaluation was sourced from a UAV Testing Competition.

\end{abstract}


\section{Introduction}
Autonomous Systems (AS) are increasingly used in sectors such as transportation, healthcare, and industrial automation \cite{jahan2019security}. Ensuring their reliability and safety is essential, as failures pose significant risks to human life and infrastructure \cite{mejri2014survey,wakabayashi2018uber}.

\begin{figure}[h]
    \centering
    \includegraphics[width=0.6\linewidth]{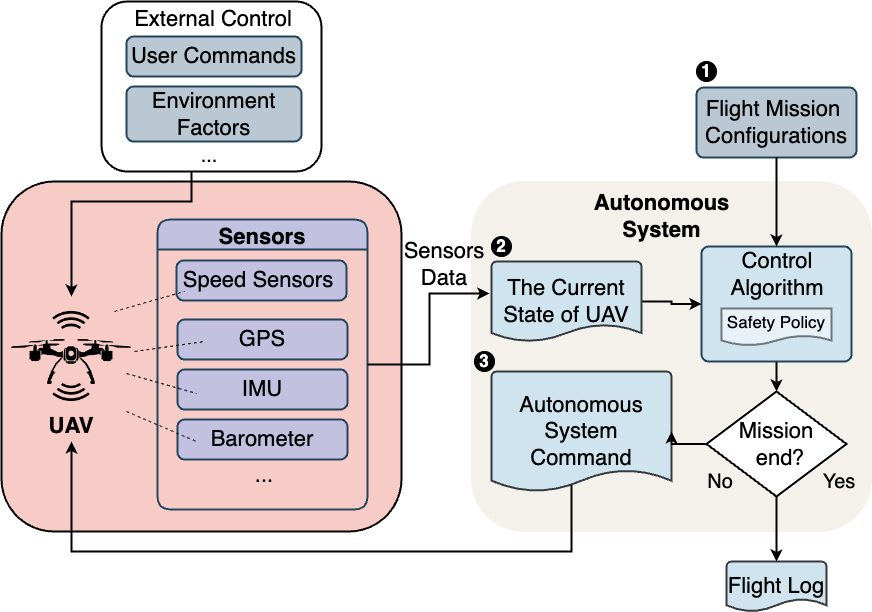}
    \caption{An Illustrative Workflow of an Autonomous System for UAVs}
    \label{fig:Autonomous System}
\end{figure}

While the testing approach proposed in this paper is applicable to general AS, we focus specifically on AS for Unmanned Aerial Vehicles (UAVs). Figure \ref{fig:Autonomous System} illustrates the functioning of such a system: \circled{1} Mission-specific configurations are provided to the control algorithm module. \circled{2} As the UAV operates, its sensors collect environmental data, forming the current state, which is relayed to the control algorithm module to calculate the next actions \cite{giering2015multi, korthals2018multi}. \circled{3} Commands are generated based on whether the mission is complete or if certain conditions hinder further progress. If the mission continues, the system outputs commands according to a predefined safety policy for the UAV to execute. The UAV then carries out these commands and sends updated sensor data back. External factors, such as wireless remote control or changing wind conditions, can also influence the UAV, and the AS adjusts accordingly based on the sensor feedback.

Fuzz testing is a software testing method that involves supplying invalid, unexpected, or random data inputs to a program in order to uncover bugs and vulnerabilities \cite{cha2015program}. This approach has been applied to fuzzing AS \cite{kim2021pgfuzz}, where seeds such as configuration parameters relevant to the safety properties under test are selected. A seed is randomly chosen for mutation, assigned random values, and input into a UAV simulator. The simulator executes the input and records whether the seed is valuable for future use, based on predefined heuristics, rather than generating entirely new random values.

Motivated by \cite{kim2021pgfuzz}, instead of relying on user-specified heuristics, we leverage the LLM's understanding of safety properties as the metric for automatically determining the relevance of a test case. In other words, our focus is on how ``interesting" a test case is in relation to specific testing objectives via LLMs. This approach shifts the emphasis of AS fuzzing to properly defining the system's safety properties and ensuring that the LLM accurately interprets these safety properties as the key objectives for testing.

For example, testers may evaluate whether a drone control system violates its hover policy by defining ``Drone altitude changes during hover mode" as an interesting test condition. Our approach transforms an LLM into a predictor that identifies which test cases are most likely to meet this condition, deeming those as interesting test cases.

In summary, this paper offers the following contributions: 1) The design of a universal fuzzing framework that integrates LLMs for testing AS, 2) The development of \toolname as a predictor for test case relevance, and 3) Its application to real-world fuzzing tools. We release \toolname at \cite{SAFLITE} for developers to test their own fuzzing tools.

\section{Approach}

We developed a framework called \framename to leverage large language models' understanding of safety properties in fuzzing tests for autonomous systems. This framework allows all autonomous systems fuzzing tools to integrate with large language models, facilitating comprehensive fuzzing tests on these systems.

\subsection{The Universal AS Fuzzing with LLMs}

\begin{figure}[h]
    \centering
    \includegraphics[width=\linewidth]{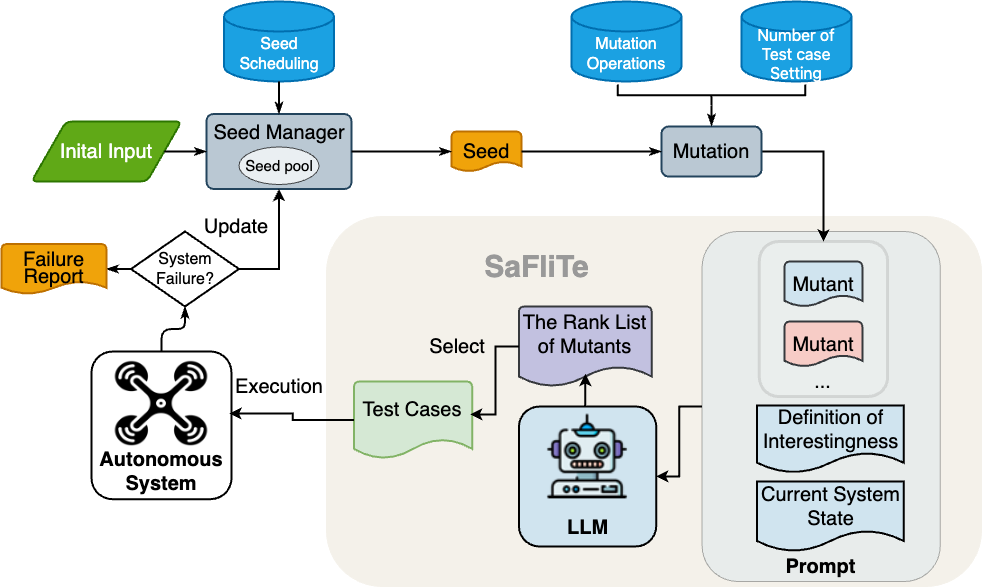}
    \caption{Workflow of the universal AS fuzzing with LLMs}
    \label{fig:interesting_fuzzing}
\end{figure}

To establish a universal fuzzing framework for AS that integrates with large language models, we initiated our approach by analysing current AS fuzzing tools including PGFuzz~\cite{kim2021pgfuzz}, DeepHyperion-UAV~\cite{zohdinasab2024deephyperion}, CAMBA~\cite{de2024camba}, and TUMB~\cite{tang2024tumb}. These tools primarily utilise mutation-based fuzzing techniques. By abstracting and refining their processes, we designed a universal AS fuzzing framework that encapsulates the methodologies of these tools. This process is shown in Figure~\ref{fig:interesting_fuzzing}, excluding the \toolname component which represents the proposed LLM integration in this study. 

The fuzzing process begins with randomly selected/generated test cases being stored as initial inputs in the seed pool within the seed manager module. The seed manager selects a test case, known as a seed, from the seed pool based on a predefined seed scheduling strategy. This seed is then transferred to the mutation module, where it is subjected to predefined mutation operations, such as randomly changing a value of parameter in the test case. This mutation creates a new test case that is executed by AS. Optionally, there could be a ``number of test case setting" in the mutation module to control the number of new test cases generated at each step by the fuzzing tool. In the event of a system failure, i.e., the violation of some safety property, a report is generated for detailed analysis. Otherwise, if the new test case does not result in a system failure, it is used by the seed manager to update the seed pool, for example, replacing the original test case with the new one, and the fuzzing process continues. 

Following the formulation of the universal AS fuzzing framework as illustrated in Figure \ref{fig:interesting_fuzzing}, we include the \toolname module prior to the AS executing the test cases, which are now first analysed through LLMs. Further details on this integration are explained in Section \ref{sec:saflite})

\begin{algorithm}
\small
\caption{The Universal AS Fuzzing with LLMs}\label{alg:interesting_fuzzing}
\begin{algorithmic}[1] 
    \State seed\_pool $\gets$ \{initial\_input\}
    \While{True}
        \State $seed \gets$ \Call{SeedManager}{initial\_pool, seed\_scheduling}
        \State $mutants \gets$ \Call{Mutation}{seed, mut\_operations, n}
        \State \highlight{test\_cases $ \gets$ \Call{\toolname}{int\_def, system\_state, mutants}}
        \If{ \Call{Execution}{test\_cases} fails}
            \State \Return system failure
        \Else
            \State update seed\_pool
        \EndIf
    \EndWhile
\end{algorithmic}
\end{algorithm}

Corresponding to the workflow shown in Figure \ref{fig:interesting_fuzzing}, the pseudocode in Algorithm \ref{alg:interesting_fuzzing} further details each step of the universal AS fuzzing framework with LLM integration. It closely aligns with the components in the workflow, where \tool{SeedManager} and \tool{Mutation} represent the seed manager module and mutation module, respectively. The parameter $n$ denotes the number of mutants generated during each mutation step. \tool{Execution} refers to the execution of these test cases by the AS, with the test cases first analysed and selected by the LLM-based module \toolname, as highlighted in the algorithm.

\subsection{\toolname }
\label{sec:saflite}
In fuzz testing, fuzzing tools for autonomous systems often produce a high volume of ineffective test cases due to the expansive input space of autonomous systems. These tools traditionally mutate seeds using predefined mutation operations, such as randomly changing the value of a parameter, which leads to inefficiency in discovering system safety issues. To address this, we have introduced an automated tool \toolname\footnote{The term \toolname{} is an abbreviation for ``\textbf{Sa}fe \textbf{Fli}ght \textbf{Te}sting'', as this paper primarily deals with fuzzing UAV autonomous systems.}, designed to improve testing efficiency without the manual review burden. \toolname employs a large language model to assess the proximity of generated test cases to predefined conditions. 
A visual depiction of \toolname's framework is provided in the highlighted box of Figure \ref{fig:interesting_fuzzing}.

\toolname adapts the LLM as its reasoning engine for fuzzing, featuring the following essential pieces:

\paragraph{Definition of Interestingness} Based on the safety specifications of an autonomous system, some test cases are deemed more ``interesting" than others due to their potential to breach safety requirements. This interestingness is defined by the expected behavior of the autonomous system when it violates these safety properties. The scope of this definition can vary from broad objectives, such as causing a control system crash, to more specific scenarios.

\begin{figure}[t]
    \centering
    \includegraphics[width=\linewidth]{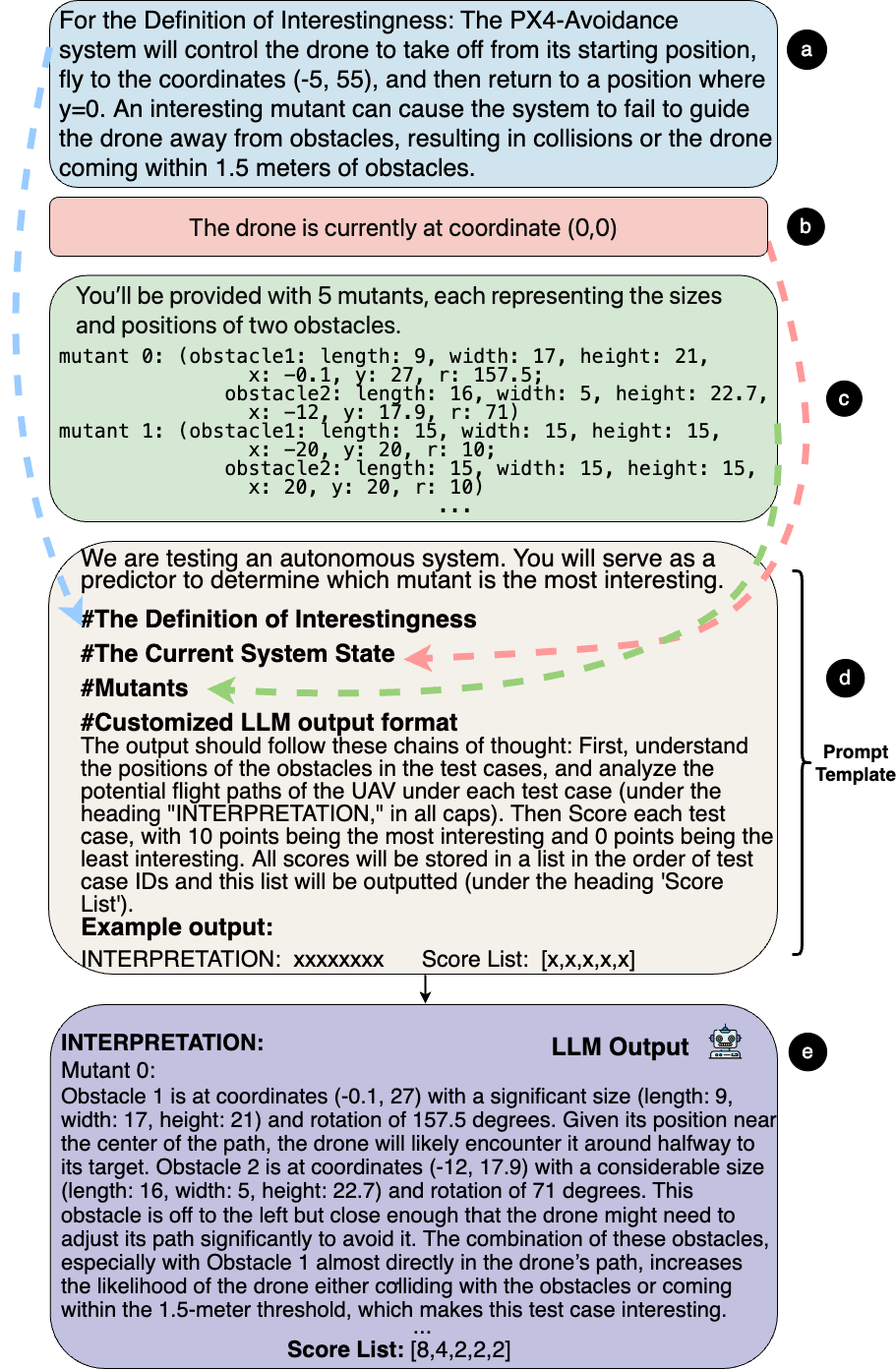}
    \caption{A concrete example of \toolname applied to a fuzzing tool testing a UAV obstacle avoidance system, PX4-Avoidance. The safety policy of system requires the UAV to maintain a distance of at least 1.5 metres from obstacles. The goal is to find test cases where the UAV either comes within 1.5 metres of an obstacle or collides with one. Each test case represents the number, size, and position of obstacles. In this example, \toolname predicts the five test cases, converting the goal into a test condition ({a}), describing the current state of UAV ({b}), and summarising the obstacles ({c}). The LLM analyses and scores each mutant using reasoning, guided by the information in ({d}).}
    \label{fig:example_ITCP}
\end{figure}

As illustrated in Figure \ref{fig:example_ITCP}\circled{a}, the definition of interestingness stipulates that the UAV must maintain a minimum distance of 1.5 meters from obstacles during the navigation of the PX4-Avoidance system's mission. This criterion emphasizes the UAV’s safe navigation, specifically designed to prevent collisions with obstacles, thus reflecting the safety requirements for operation. This example is from the UAV Testing Competition \cite{khatiri2024sbft}.

\paragraph{Current System State} It refers to the condition of the autonomous system immediately before \toolname analyses the test cases. 
For example, when testing a UAV control system, the current system state might include details such as the current location of drone and additional parameters like GPS coordinates or pitch angle. The selection of specific information to include can be informed by the definition of interestingness, which interprets the safety specifications, aiming for improving the accuracy of the assessment regarding how closely the system aligns with the predefined test conditions.

The example in Figure \ref{fig:example_ITCP}\circled{b} shows the coordinates of the UAV under testing, which represent the current state information essential for the navigation task.

\paragraph{Mutants} These represent the outcomes of mutations performed by the fuzzing tool. It is helpful to include a more detailed description of these results when presenting them, as this will assist the LLM in accurately understanding each test case.

Figure \ref{fig:example_ITCP}\circled{c} provides a concrete example of constructing a Mutants piece. Information from five mutants is extracted and formatted in a structured way, informing the LLM that each mutant represents the sizes and positions of two obstacles.

\paragraph{Prompt Template} To enable \toolname to work effectively with different LLMs as the prediction core while ensuring consistent output, we developed a prompt template that systematically incorporates all relevant information to be provided to the LLM and standardises its output. As shown in Figure \ref{fig:example_ITCP}, the template (\circled{d}) is populated using the definition of interestingness (\circled{a}), the current system state (\circled{b}), and the corresponding mutants (\circled{c}). 

Moverover, \toolname employs the CoT (Chain of Thoughts) approach to customise the LLM output format. The LLM should interpret the meaning of each mutant, then evaluate how each mutant will impact the current system state, based on the information provided about the current state. The LLM should produce a brief explanation for each thought process and assign a score to each mutant. A score of 10 indicates that the test case is the most interesting, meaning it is highly likely to meet the test conditions outlined in the definition of interestingness after execution, while a score of 0 signifies the opposite. 

The final part of the prompt template is an \emph{Example Output}, which helps ensure the LLM consistently formats its responses. When \toolname is applied to the AS fuzzing tools (as shown in Figure \ref{fig:interesting_fuzzing}), the default expected output is a selection of test cases from a ranked list of mutants.

In addition, the LLM provides insights for each test case in its output, upon which it generates a final ranked list of test case scores. As illustrated in Figure \ref{fig:example_ITCP}\circled{e}, the LLM assesses obstacle positions in the environment and evaluates how proximity to waypoints might impact the avoidance system of UAV.

\begin{algorithm}
\small
\caption{\toolname}\label{alg:saflite}
\begin{algorithmic}[1] 
    \State $prompt \gets$ \Call{SetPrompt}{int\_def, system\_state, mutants}
    \State $rank\_ist \gets$ \Call{LLMAgent}{prompt} 
    \State test\_cases $ \gets$ \Call{Select}{rank\_list}
    \State \Return test\_cases
\end{algorithmic}
\end{algorithm}

In summary, the pseudocode in Algorithm \ref{alg:saflite} implements the \toolname component of Algorithm \ref{alg:interesting_fuzzing}. The execution process of \toolname begins with \tool{SetPrompt}, which automatically populates the prompt template with the three required inputs: the definition of interestingness ($int\_def$), the current system state ($system\_state$), and the mutants to be analysed ($mutants$). Next, \tool{LLMAgent} interacts with the LLM using this prompt to generate a ranked list of test cases. Finally, \tool{Select} defines the strategy for selecting test cases from the ranked list that are most closely aligned with the expected test condition in the autonomous system, such as setting thresholds for interestingness and uninterestingness values.

\section{Evaluation}

In this section, we evaluate the proposed \toolname approach by addressing three research questions (RQs): validating that LLMs can indeed identify interesting test cases (Section \ref{sec:rq1}), showing that a more specific definition of interestingness can further improve its effectiveness (Section \ref{sec:rq2}), and, utmost, confirming that \toolname significantly improves the performance of existing fuzzing tools (Section \ref{sec:rq3}).

\subsection{RQ1: How effective do LLMs predict the interestingness of test cases?} 
\label{sec:rq1}

\paragraph{{Experimental Setup}} For RQ1, we provided \toolname with a labelled dataset for prediction and evaluation. The dataset comprises 117 log files representing both flight issues (interesting) and normal flights (non-interesting), sourced from a combination of real-world and synthesised flights \cite{PX4,khatiri2024sbft}. The interesting logs were drawn either from the PX4 GitHub issue tracker, where users had uploaded logs that were accepted as demonstrating buggy behaviour requiring code changes, or from simulated flights replicating similar issues. The non-interesting logs were sourced from two places: simulated flights conforming to a standardised test card used by PX4 developers to validate code changes, and flights completed during the development of a new navigation submodule. These non-interesting logs showed standard mission completions and adhered to the policy requirements of the additional submodule. We broadly define interestingness as the occurrence of uncommon, difficult-to-understand, or significantly unexpected behaviors in drones, which may potentially lead to high-risk flights.

To assess the effectiveness of \toolname across different LLMs, we selected three models: GPT-3.5 \cite{ouyang2022training}, Mistral-7B \cite{jiang2023mistral}, and Llama-2-7B \cite{touvron2023llama} such that Mistral-7B and Llama-2-7B are deployed locally. The evaluation was conducted using four metrics: Accuracy ($Acc$), Precision, ($Prec$) Recall, and F1-score ($F1$). 

\begin{table}[ht]
\centering
\resizebox{\linewidth}{!}{
    \begin{small}
    \begin{tabular}{ lrrrr } 
    \toprule
    LLMs & Acc & F1 & Prec & Recall  \\
    \midrule
    LLAMA-2-7B & 47.1\% & 54.3\% & 48.9\% & 61.1\%   \\
    Mistral-7B & 62.9\% & 66.7\% & 61.9\% & \cellcolor{cyan!20}\textbf{72.2\%}   \\
    GPT-3.5 & \cellcolor{cyan!20}\textbf{68.6\%} & \cellcolor{cyan!20}\textbf{68.6\%} & \cellcolor{cyan!20}\textbf{70.6\%} & 66.7\%   \\
    GPT-3.5 (temp=1) & 52.4\% & 46.4\% & 55.9\% & 39.8\%   \\
    \bottomrule
    \end{tabular}
    \end{small}
}
\caption{Performance comparison of different LLMs for predicting the interestingness of flight log benchmarks.}
\label{tab:different_llms}
\end{table}

\paragraph{{Results}} The results for the four metrics are summarised in Table \ref{tab:different_llms}, showing that GPT-3.5 achieves the highest performance in three metrics, with an accuracy of 68.6\% and an F1 score of 68.6\%. Its precision of 70.6\% underscores GPT-3.5's strong ability to identify interesting test cases. The overall accuracy metric indicates that Llama-2-7B, the local LLM, performs the least effectively. However, another local LLM, Mistral-7B, performs comparably to GPT-3.5 and even surpasses it in recall, suggesting a strong capability to distinguish between interesting and non-interesting test cases. This demonstrates that smaller, locally deployed LLMs can achieve performance comparable to GPT-3.5. Given the advantages of local LLMs for secure, local deployments, task-specific fine-tuning of Mistral-7B could further enhance its potential as the reasoning engine for \toolname.

An analysis of the recall metric reveals that GPT-3.5's recall drops significantly when its temperature is set to 1, indicating confusion in distinguishing between interesting and non-interesting test cases. This suggests that a high temperature results in overly random outputs and fabricated responses without adequate reasoning. Therefore, we recommend avoiding high-temperature settings when using \toolname.

\begin{tcolorbox}[colback=gray!20, colframe=black, boxrule=0.2mm, rounded corners, boxsep=0.1mm]
Answer to RQ1: LLMs, such as GPT-3.5 and Mistral-7B, demonstrate effectiveness in predicting the interestingness of the flight log benchmark, though their performance is not exceptional. The broad definition of interestingness likely plays a key role in limiting their effectiveness.
\end{tcolorbox}

\subsection{RQ2: How does a more specific definition of interestingness impact the effectiveness of \toolname?}
\label{sec:rq2}

\paragraph{{Experimental Setup}} To examine the impact of using a more specific definition of interestingness on the effectiveness of \toolname, we conducted experiments by integrating \toolname with the existing fuzzing tool PGFuzz \cite{kim2021pgfuzz}.

PGFuzz can perform fuzzing tests on ArduPilot \cite{arduPilot} and PX4 \cite{PX4} systems based on safety policies to detect policy violations.  We obtained a list of bugs from the PGFuzz repository\footnote{The bug list is available in Section 4 of the README on the PGFuzz GitHub repository at \url{https://github.com/purseclab/PGFuzz}.}. Since most bugs in the list lacked details about the system state or violated policies, we selected 8 bugs for our experiment. For each bug, we translated the associated policy violation into a natural language description, which served as a definition of interestingness for \toolname. We then provided \toolname with the control system's state at the time of the bug occurrence (details in APPENDIX \ref{appendix:A}). Given that PGFuzz randomly selects a parameter from a list related to the target policy as a test case, we extracted all possible parameters from this list and provided them to \toolname for analysis. Each test case was scored: below 4 as non-interesting, 5 to 7 as mid-interesting, and above 7 as interesting. If the test case that caused the bug was rated mid-interesting or interesting, it indicates that \toolname improves the likelihood of selecting bug-causing test cases over PGFuzz’s random selection, thereby enhancing efficiency and demonstrating the value of using more specific Definitions of Interestingness.

\begin{figure}
    \centering
    \includegraphics[width=\linewidth]{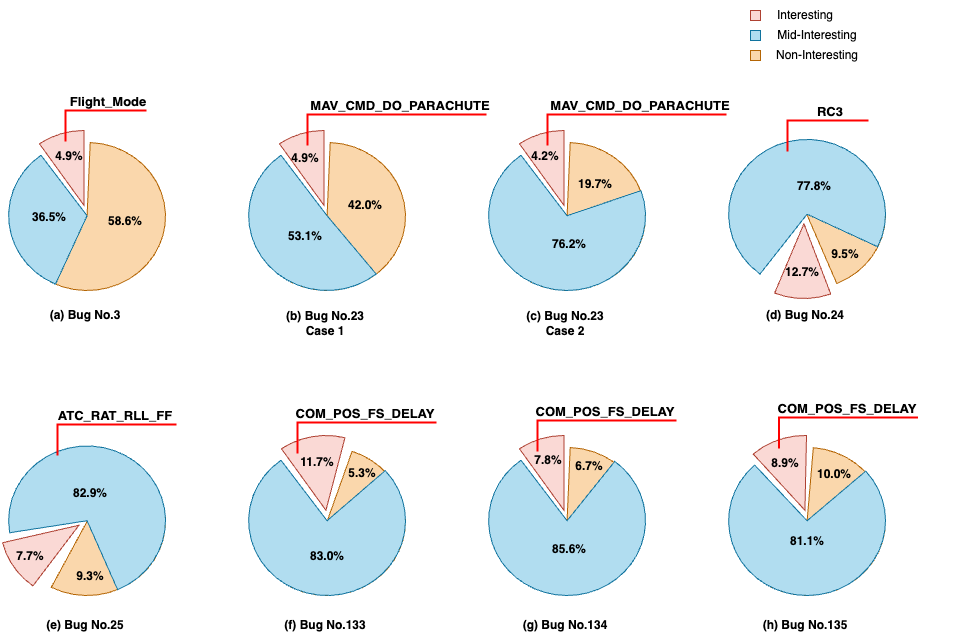}
    \caption{For each bug, \toolname categorizes all potential test cases generated by PGFuzz into three groups based on the corresponding Current State and Definition of Interestingness. The figure highlights which category the test case that triggered the bug was assigned to.}
    \label{fig:bug_analyze}
\end{figure}

\paragraph{{Results}} The results of this experiment are presented in Figure \ref{fig:bug_analyze}. Each pie chart represents the outcomes of using \toolname to analyse a specific violated safety property. For example, Figure 4a shows that, for the selected safety property and across all test cases generated by PGFuzz, \toolname classified 58.8\% of test cases as non-interesting, 36.5\% as mid-interesting, and only 4.9\% as interesting. Although the proportion of interesting test cases is the smallest, the test case involving \tool{Flight\_Mode}, which caused the bug, was accurately classified as interesting. Therefore, compared to PGFuzz's random selection of all possible test cases, using \toolname to select only from the interesting category increases the likelihood of selecting a test case that leads to a policy violation by 93.1\% in each fuzzing round. Overall, \toolname effectively classified most test cases that resulted in policy violations as interesting, with only two classified as mid-interesting and none as non-interesting. On average, the probability of selecting a bug-causing test case in each fuzzing round increased by 94.7\%.

Additionally, we observed that GPT-3.5 accurately understands the function of each test case and its impact on drone control systems. For example, GPT-3.5 correctly identified the test case \tool{MAV\_CMD\_DO\_PARACHUTE} as a command to deploy the UAV parachute (see APPENDIX \ref{appendix:B} for further test case analyses). This suggests that GPT-3.5's training data might include considerable information about ArduPilot and PX4 systems. However, the test cases that caused bugs No.24 and No.25 were categorised as mid-interesting instead of Interesting. Upon review, this occurred because GPT-3.5 lacked knowledge of the specific values assigned to these test cases, making it difficult to predict their exact impact (details in APPENDIX \ref{appendix:B}).
As correct classification is challenging without specific values, this demonstrates that GPT-3.5 recognises that some value assignments may not cause bugs, while others might. This underscores the importance of retaining some randomness in fuzzing when testing automated systems and highlights the value of integrating \toolname with fuzzing tools like PGFuzz.

\paragraph{New bugs} Both PGFuzz and PGFuzz+\toolname were given 24 hours to perform fuzzing tests on each policy. The results were compared based on the number of bugs discovered and the number of fuzzing rounds completed. Notably, when testing the ``PX.HOLD1" policy, both PGFuzz and PGFuzz+\toolname identified the same policy violation. However, PGFuzz+\toolname achieved this in just 126 rounds of fuzzing, significantly fewer than the 269 rounds required by PGFuzz.  Additionally, the bug identified by PGFuzz+\toolname is a new bug in the PX4 system that had not been previously identified and was absent from the bug list presented by PGFuzz. The bug occurred as follows: \circled{1} The drone was initially in Hold mode, hovering in place. \circled{2} \toolname analysed the test cases, and the {MAV\_CMD\_DO\_REPOSITION} test case was assigned a score of 7. \circled{3} When the PX4 system executed the {MAV\_CMD\_DO\_REPOSITION} test case, the drone began moving northwest but remained in Hold mode.

Moreover, during the study, we also discovered a new bug. The bug occurred as follows: \circled{1} The PX4 system was preparing to switch to ``ACRO" flight mode during a mission. \circled{2} \toolname analyzed the test cases and gave a score of 8 to the {MAV\_CMD\_DO\_PARACHUTE} test case. \circled{3} The PX4 system switched to ``ACRO" flight mode and executed {MAV\_CMD\_DO\_PARACHUTE}. Although the safety policy prohibits deploying the parachute in ``ACRO" mode, the parachute was deployed, causing the drone to crash.

\begin{tcolorbox}[colback=gray!20, colframe=black, boxrule=0.2mm, rounded corners, boxsep=0.1mm]
Answer to RQ2: Using \toolname with a specific definition of interestingness as the test condition can enhance the effectiveness of existing fuzzing tools, particularly by improving their ability to understand and predict test cases. 
\end{tcolorbox}

\subsection{RQ3: How effectively does \toolname improve the performance of existing AS fuzzing tools?}
\label{sec:rq3}

\paragraph{{Experimental Setup}} To assess whether \toolname can successfully integrate with fuzzing tools and improve their performance in a full fuzzing test process, we followed the structure of \framename and combined \toolname with fuzzing tools from the SBFT UAV Testing Competition \cite{khatiri2024sbft}: DeepHyperion-UAV \cite{zohdinasab2024deephyperion}, TUMB \cite{tang2024tumb}, and CAMBA \cite{de2024camba}. The competition tasked these fuzzing tools with testing the PX4-Avoidance system, where a UAV navigates missions in a virtual environment under the control of PX4-Avoidance. The objective was to generate as many valid test cases as possible. A test case defines the size and coordinates of obstacles in the environment and is considered valid if the UAV either collides with an obstacle or comes too close. We translated this valid test case rule into the definition of interestingness for \toolname, as in Figure \ref{fig:example_ITCP}\circled{a}.

In our experiments, we adhered strictly to the competition rules and ran all fuzzing tools in the designated environment. Since automated control systems may behave differently across various missions, we provided six distinct missions (Mission 2-7) to compare the performance of the fuzzing tools in different scenarios. Mission 1 was excluded, as it was intended for participants to test their tools during development. Each mission, defined by the competition organisers, varied only in specific waypoints, requiring the system to guide the drone through them in sequence. We compared the number of valid test cases generated by the modified tools (integrated with \toolname) to those produced by the original tools used in the competition.

\paragraph{{Results}} The experimental results, presented in Table \ref{tab:CPS-UAV_comparison}, demonstrate that \toolname successfully integrates with various fuzzing tools, completing the full fuzzing process and improving their performance. Each integrated tool generated more valid test, with improvements observed in all the missions.

Among the three tools, DeepHyperion-UAV exhibited the most significant improvement in the number of valid test cases. However, the gains for CAMBA and TUMB were more modest. Upon reviewing their code, we believe this is due to their approach of making only minor mutations to each test case when generating new ones. As a result, the test cases produced are often very similar, and the LLM assigns similar scores, reducing the ability of the tools to rank test cases effectively based on their relevance to the Definition of Interestingness.

\begin{table}[ht]
\centering
\resizebox{\linewidth}{!}{
    \begin{large}
    \begin{tabular}{ lrrrrrr } 
    \toprule
     & \multicolumn{6}{c}{\textbf{Number of Valid Test Cases}}  \\
     \cmidrule(lr){2-7}
    \textbf{Tool Name} & \textbf{Mission 2} & \textbf{Mission 3} & \textbf{Mission 4} & \textbf{Mission 5} & \textbf{Mission 6} & \textbf{Mission 7}  \\
    \midrule
    DeepHyperion-UAV & 2 & 10 & 8 & 4 & 0 & 5  \\
    DeepHyperion-UAV + \textbf{\toolname} & \textbf{17} & \textbf{15} & 8 & \textbf{10} & \textbf{25} & \textbf{22} \\
    \toprule
    CAMBA & 7 & 9 & 0 & 1 & 6 & 10 \\
    CAMBA + \textbf{\toolname} & \textbf{11} & \textbf{11} & \textbf{2} & \textbf{3} & 6 & \textbf{11} \\
    \toprule
    TUMB & 2 & 15 & 11 & 7 & 18 & 20 \\
    TUMB + \textbf{\toolname} & \textbf{5} & 15 & \textbf{18} & 7 & \textbf{20} & \textbf{21}  \\
    \bottomrule
    \end{tabular}
    \end{large}
}
\caption{Comparison of the number of valid test cases generated by DeepHyperion-UAV, CAMBA, and TUMB in on the UAV Testing Competition benchmark before and after the integration of \toolname. Each valid test case represents a unique scenario that causes the UAV mission to fail.}
\label{tab:CPS-UAV_comparison}
\end{table}

Notably, all three fuzzing tools integrated with \toolname showed significant improvements in the number of valid test cases generated in Mission 2. In the other missions, at least one tool either showed no increase or achieved only a single additional valid test case. This variability in performance across missions may be attributed to the complexity of the tasks. Mission 2, being the simplest, involves the drone navigating to a waypoint and returning, which could explain the more consistent performance. However, as the missions become more complex, the LLM may struggle to fully understand the mission, leading to unstable test case predictions and inconsistent performance. To address this, fine-tuning the LLM for specific missions or providing more detailed descriptions of the mission in the Current System State within the prompt might help improve its understanding and improve the accuracy of test case predictions.

\begin{table}[!htp]
\centering
\resizebox{\linewidth}{!}{
    \begin{large}
    \begin{tabular}{ lrrrrrr } 
    \toprule
     & \multicolumn{6}{c}{\textbf{Number of Valid Test Cases}}  \\
     \cmidrule(lr){2-7}
    \textbf{LLM} & \textbf{Mission 2} & \textbf{Mission 3} & \textbf{Mission 4} & \textbf{Mission 5} & \textbf{Mission 6} & \textbf{Mission 7}  \\
    \midrule
    GPT3.5 & 17 & 15 & 8 & 10 & 25 & 22  \\
    Mistral-7B & 13 & 11 & 13 & 8 & 33 & 14 \\
    Llama-2-7B & 12 & 10 & 8 & 4 & 16 & 8 \\
    \bottomrule
    \end{tabular}
    \end{large}
}
\caption{Results of the number of valid test cases generated by DeepHyperion-UAV + \toolname using different LLMs.}
\label{tab:CPS-UAV different_llms}
\end{table}

\paragraph{Different LLMs} We further investigate the outcomes from DeepHyperion-UAV+\toolname, using different LLMs as the prediction core. As shown in Table \ref{tab:CPS-UAV different_llms}, overall, GPT-3.5 still delivers the most significant improvement in the capabilities of DeepHyperion-UAV, while Llama-2-7B offers relatively less improvement. Notably, Mistral-7B demonstrates performance improvements that are nearly on par with GPT-3.5 and even surpasses GPT-3.5 in Mission 4 and Mission 6. The performance rankings of the three LLMs in fuzzing align with the results from RQ1, with GPT-3.5 leading, Mistral-7B close behind, and Llama-2-7B showing the weakest performance.

Given the results, Mistral-7B's strong performance is particularly impressive, considering it is a local LLM that has not undergone fine-tuning. considering it is a local LLM that has not undergone fine-tuning. This highlights the suitability of locally deployed, smaller LLMs for AS fuzzing tasks, and further improvements could unlock even greater potential for such models in this domain.

\begin{tcolorbox}[colback=gray!20, colframe=black, boxrule=0.2mm, rounded corners, boxsep=0.1mm]
Answer to RQ3: \toolname can successfully integrate with various fuzzing tools, execute a full fuzzing test, and significantly improve their performance.
\end{tcolorbox}

\section{Related Work}
In fuzzing AS, most tools rely on mutation-based techniques. For example, Hu et al. mutated traffic scenarios by changing traffic light colors or vehicle starting positions for testing the autonomous driving systems\cite{hu2021coverage}. Additionally, there is also a lot of research focused on fuzzing UAV AS. For example, Khatiri et al. organized a fuzzing competition for the PX4-avoidance system, aiming to generate test cases that cause policy violations \cite{khatiri2024sbft}. This led to the development of several fuzzing tools for UAV systems. Zohdinasab et al. introduced a novel algorithm to rank seeds, selecting the best ones for mutation and test case generation \cite{zohdinasab2024deephyperion}. Tang et al. developed TUMB, which uses Monte Carlo Tree Search (MCTS) to explore the UAV environment and generate diverse test cases \cite{tang2024tumb}. Similarly, Winsten et al. \cite{winsten2024adaptive} applied Wasserstein generative adversarial networks \cite{arjovsky2017wasserstein} to generate UAV test cases, comparable to Zhong et al.'s use of neural networks in fuzzing autonomous driving systems to predict whether new seeds would trigger traffic violations \cite{zhong2022neural}. Both approaches exploit neural networks' ability to manage complex relationships. However, training neural networks can pose significant stability and convergence issues \cite{pascanu2013difficulty}. In contrast, we leverage LLMs to predict test cases, offering a more adaptive approach. Current research on LLMs in AS focuses on environmental understanding. For instance, Jiahui et al. showed that LLMs can assess the realism of driving environments \cite{wu2024reality}. Similarly, PromptTrack, introduced by Dongming et al., enables LLMs to predict object trajectories \cite{wu2023language}, while HiLM-D predicts risk object locations from images \cite{ding2023hilm}. Our research differs by focusing on using LLMs to predict interesting test cases based on their environmental understanding.

\section{CONCLUSION}
In this paper, we introduce \toolname, a predictor that uses an LLM to determine whether a test case aligns with a user-defined test condition. We designed a framework for \toolname that integrates with any fuzz testing tool for AS, using its predictive capabilities to filter out irrelevant test cases and enhance performance. We evaluated the prediction accuracy of \toolname using real flight logs and tested it with four fuzz testing tools on three automated systems. The results show that \toolname significantly improves fuzz testing efficiency.

\bibliographystyle{ieeetr}
\bibliography{all}

\appendix
\subsection{Bugs Information Provided to \toolname}\label{appendix:A}
We selected information related to 8 bugs, as shown in Table \ref{tab:Bugs_Information}, and provided it to \toolname to evaluate its performance.
\begin{table}[h]
\centering
\tiny
\begin{tabular}{|m{1cm}<{\centering}|m{2.2cm}|m{4cm}|}
\hline
\thead{ID} & \thead{Current State} & \thead{Definition of Interestingness} \\
\hline
Bug NO.3 & The UAV is currently at an altitude of 20 meters, and the flight mode has switched from MISSION to FLIP. & \vspace{1pt}Policy: If and only if roll is less than 45 degree, throttle is greater or equal to 1,500, altitude is more than 10 meters, and the current flight mode is one of ACRO and ALT\_HOLD, then the flight mode can be changed to FLIP.
If a test case has the potential to cause the policy violation or drone crash, then this test case is considered an interesting test case. \vspace{1pt}\\
\hline
Bug NO.23 case 1 & Flight Mode: FLIP & \vspace{1pt}Policy: Deploying a parachute requires following conditions: (1) the motors must be armed, (2) the vehicle must not be in the FLIP or ACRO flight modes, (3) the barometer must show that the vehicle is not climbing, and (4) the vehicle’s current altitude must be above the CHUTE\_ALT\_MIN parameter value. If a test case has the potential to cause the policy violation or drone crash, then this test case is considered an interesting test case. \\
\hline
\vspace{1pt}Bug NO.23 case 2 \vspace{1pt}& \vspace{1pt}Flight Mode: MISSION; RC 3 1900 \vspace{1pt}& It is the same as Bug NO.23 case 1 \\
\hline
Bug NO.24 & \vspace{1pt}The value of the GPS\_POS1\_Z parameter is 2.474742, causing a significant deviation in the GPS sensor measurements. The flight mode has switched from FLIP to ALT\_HOLD. \vspace{1pt}& Policy:If the Mode is ALT\_HOLD and the throttle stick is in the middle (i.e., 1,500) the vehicle must maintain the current altitude. If a test case has the potential to cause the policy violation or drone crash, then this test case is considered an interesting test case. \\
\hline
Bug NO.25 & \vspace{1pt}Flight Mode: ALT\_HOLD, Throttle: 1500 & It is the same as Bug NO.24  \\
\hline
Bug NO.133 & Flight Mode: ORBIT. After the test case is executed, the GPS will be turned off. & \vspace{1pt}Policy:If time exceeds COM\_POS\_FS\_DELAY seconds after GPS loss is detected, the GPS fail-safe must be triggered. If a test case has the potential to cause the policy violation or drone crash, then this test case is considered an interesting test case. \\
\hline
Bug NO.134 & \vspace{1pt}Flight Mode: Mission. Flying into another location. After the test case is executed, the GPS will be turned off. & It is the same as Bug NO.133 \\
\hline
Bug NO.135 & \vspace{1pt}Flight Mode: Mission. Just took off. After the test case is executed, the GPS will be turned off. & It is the same as Bug NO.133 \\
\hline
\end{tabular}
\caption{The Current State and the Definition of Interestingness extracted from the bug list in the PGFuzz repository.}
\label{tab:Bugs_Information}
\end{table}

\subsection{The Analysis Results of \toolname}\label{appendix:B}
Table \ref{tab:analysis_results} shows the analysis results of \toolname for test cases that lead to policy violations.
\begin{table}[h]
\centering
\tiny
\begin{tabular}{|m{2.2cm}<{\centering}|m{6cm}|}
\hline
\thead{Test Case} & \thead{\toolname Analysis Result} \\
\hline
Flight\_Mode & \vspace{2pt}INTERPRETATION: The test case represents setting the flight mode to 2.
it could potentially violate the policy as it is not one of the approved modes (ACRO or ALT\_HOLD) for the FLIP operation. This could lead to the drone performing a flip maneuver in an unsafe or unexpected state, potentially causing a crash or loss of control. Score: 8 out of 10. \vspace{2pt}\\
\hline
MAV\_CMD\_DO\_PARACHUTE & \vspace{2pt}INTERPRETATION: The test case represents a user command to deploy the parachute. This test case is highly interesting as deploying a parachute is a critical safety measure. It could potentially cause a policy violation or a drone crash if the conditions for deploying a parachute are not met, such as the motors not being armed, being in the wrong flight mode, or the vehicle not meeting altitude and climbing conditions. Therefore, any command related to parachute deployment requires careful consideration to ensure the safety and integrity of the UAV. Score: 10 out of 10. \vspace{2pt} \\
\hline
RC3 & \vspace{2pt}INTERPRETATION: The test case represents a user input to the control system, possibly related to the remote control channel 3. This test case is quite interesting because it involves user input that could potentially affect the mode or throttle stick position, which could lead to a policy violation or drone crash. The specific impact would depend on the exact nature of the RC3 command and the current state of the UAV. Score: 7 out of 10.\vspace{2pt} \\
\hline
ATC\_RAT\_RLL\_FF & \vspace{2pt}INTERPRETATION: This test case likely represents a command related to the roll rate feedforward for the aircraft's control system. This test case is moderately interesting because it involves a parameter related to the control system. Changes in the roll rate feedforward could potentially affect the stability and control of the UAV, which may lead to a policy violation or drone crash if not handled properly. However, without specific values or context, it's difficult to determine the exact impact of this command. Score: 5 out of 10. \vspace{2pt} \\
\hline
COM\_POS\_FS\_DELAY & \vspace{2pt}INTERPRETATION: This test case represents the delay time after GPS loss is detected before the GPS fail-safe is triggered. This test case is quite interesting because it directly relates to the policy of triggering GPS fail-safe after a certain delay. If the value assigned to COM\_POS\_FS\_DELAY is too high, it could potentially lead to a policy violation or drone crash if the fail-safe is not triggered in time after GPS loss is detected. Therefore, it is important to carefully analyze and test this parameter to ensure the safety and compliance with the policy. Score: 8 out of 10. \vspace{2pt} \\
\hline
\end{tabular}
\caption{\toolname provided its interpretation and scoring of the test cases.}
\label{tab:analysis_results}
\end{table}

\end{document}